\documentclass{moriond}

\usepackage[T1]{fontenc} % if needed
\usepackage{amsmath}
\usepackage{graphicx}
\usepackage{epstopdf}
\usepackage{bm}
\usepackage{epsfig}
\usepackage{graphics}
\usepackage{xspace}
\usepackage{hyperref}
\usepackage{slashed}
\usepackage{xcolor}
\usepackage{longtable}
\usepackage[small]{subfigure}
\usepackage[normalem]{ulem}
\usepackage[symbol]{footmisc}

\def\be{\begin{equation}}
\def\ee{\end{equation}}
\def\bea{\begin{eqnarray}}
\def\eea{\end{eqnarray}}

\newcommand{\Photo}{\centering \includegraphics[height=35mm]{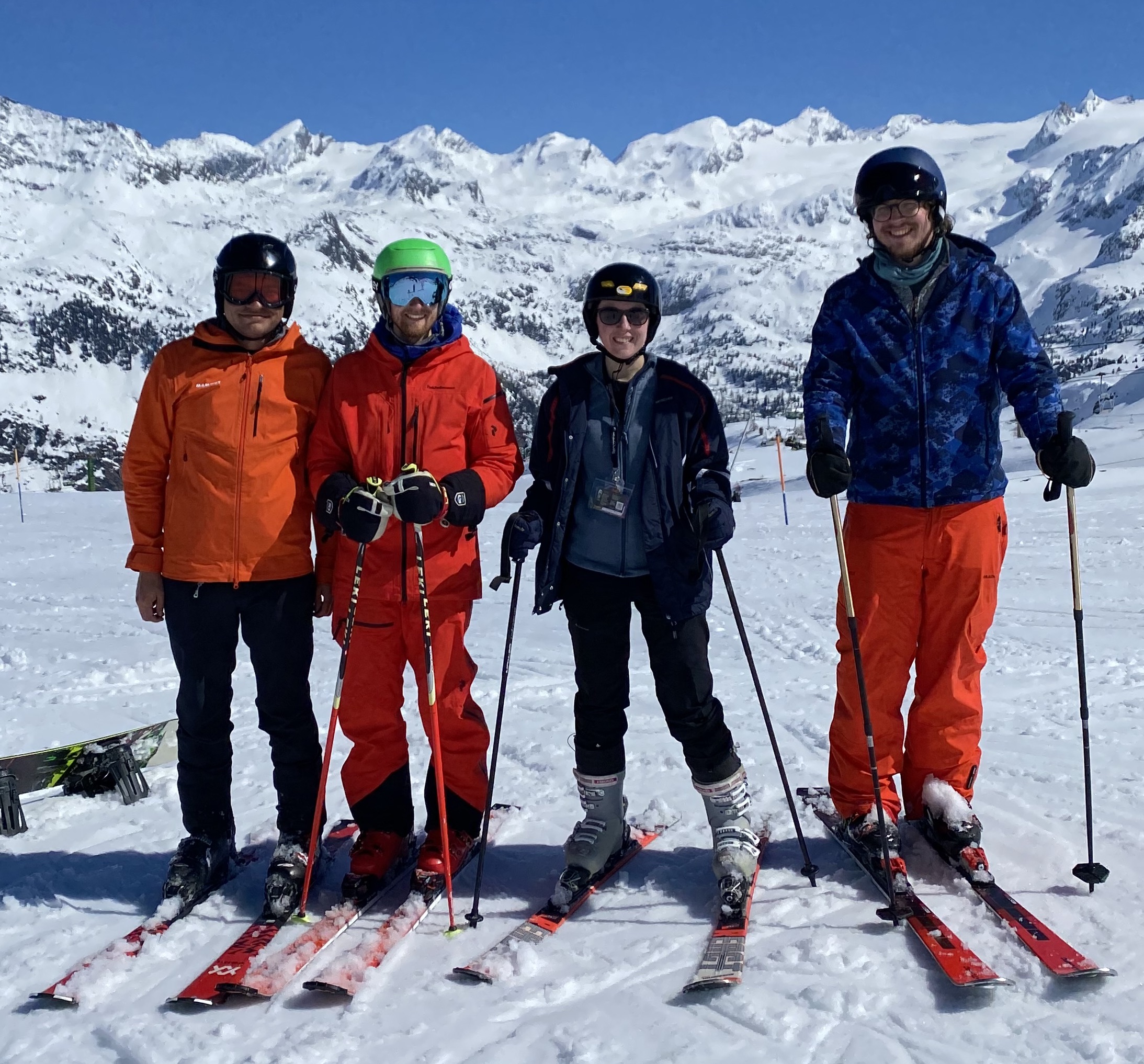}}

\begin{document}

\vspace*{4cm}
\title{Determining $\alpha_s(m_Z)$ from the Heavy Jet Mass distribution}

\author{Miguel A. Benitez}

\address{\flushleft \quad Dpto.\ de F\'isica Fundamental e IUFFyM, Universidad de Salamanca, E-37008 Salamanca, Spain}

\maketitle\abstracts{
We present a state-of-the-art analysis on determining the strong coupling constant $\alpha_s(m_Z)$ from available $e^+e^-$ data for the Heavy Jet Mass (HJM) distribution. Dijet resummation is supplemented with additional resummation of shoulder logarithms appearing around the symmetric 3-jet configuration of the HJM spectrum. In addition to $\alpha_s(m_Z)$ we further obtain numerical results for two non-perturbative parameters, $\Omega_1^{\rho}$ and $\Theta_1$, encompassing effects from dijet and trijet hadronization effects, respectively.}

\vspace*{-1cm}

\section{Introduction}\label{sec:Intro}

Determinations of the strong coupling constant from available $e^+e^-$ data for the HJM distribution have, in the past, yielded results below similar determinations from more inclusive observables like Thrust or the C-parameter~\cite{Salam:2001bd,Dissertori:2009ik,Chien:2010kc}. In recent years, a number of differences between HJM and the aforementioned event shapes have been analyzed. In particular, resummation of logarithms related to the left-Sudakov shoulder, which appear when approaching $\rho \to 1/3$ from the left, has been studied~\cite{Bhattacharya:2022dtm,Bhattacharya:2023qet}. Furthermore, Refs.~\cite{Caola:2021kzt,Caola:2022vea} found evidence for a negative non-perturbative power correction for HJM when having three jets in the underlying kinematic configuration. We present a thorough phenomenological analysis, touching upon these and other theoretical developments and their impact in the determination of the strong coupling constant.

\section{Theoretical Description}\label{sec:theory}

Our theoretical description of the HJM distribution consists of two main parts. In the dijet limit, the differential cross section factorizes into a hard function, two jet functions and a soft function~\cite{Catani:1991bd,Korchemsky:1999kt,Schwartz:2007ib}
\begin{equation}\label{eq:Xsec}
%\frac{1}{\sigma_0}\frac{\df \sigma}{\rm d \rho} = %2\,Q^2\,
\text{d} \sigma_{\text{dij}} = 
H_{\text{dij}}\times J_1 \times J_2\otimes S_{1,2}
\otimes F^{\Xi}_{1,2}(\Omega_1^{\rho})\,.
\end{equation}
$F^{\Xi}_{1,2}(\Omega_1^{\rho})$ denotes the shape function encoding non-perturbative corrections, which are formally manifested by matrix elements of QCD operators. Around the symmetric 3-jet limit $\rho \to 1/3$, the distribution factorizes as~\cite{Bhattacharya:2022dtm,Bhattacharya:2023qet}
\begin{equation}\label{eq:Xsecsh}
% \frac{1}{\sigma_0}\frac{\df \sigma}{{\rm d} r} =
 \text{d}\sigma_{\text{sh}}^{\text{pert}} = H_{\text{sh}}\times J_1\times J_2 \times J_3 \otimes S_{1,2,3}\,.
\end{equation}
In Eq.~(\ref{eq:Xsecsh}), non-perturbative corrections enter through a shift of the perturbative partonic cross section
\begin{equation}
     \frac{\text{d}\sigma_{\text{sh}}}{\text{d} \rho}(\rho)
     =\frac{\text{d}\sigma^\text{pert}_{\text{sh}}}{\text{d} \rho}\biggl(\rho -\frac{\Theta_1}{Q}\biggr)\,,
\end{equation}
where the non-perturbative shift parameter $\Theta_1$ conceptually encodes information on the hadronization effects coming from a kinematic configuration with three jets in the final state. The matching between the dijet, shoulder and fixed-order regions can subsequently be written as
\begin{equation}
    \text{d}\sigma = \Bigl[\text{d}\sigma_{\text{dij}} - \text{d}\sigma_{\text{dij}}^{\text{sing}}\Bigr] + \text{d}\sigma_{\text{FO}}
    + \Bigl[\text{d}\sigma_{\text{sh}} - \text{d}\sigma_{\text{sh}}^{\text{sing}}\Bigr]\,.
\end{equation}
In practice, analogous to the case of dijet resummation, the interpolation between regions is handled through suitable sets of profile functions. They encode the evolution of the jet and soft scales as a function of $\rho$ in the dijet, as well as the shoulder region.

\section{Experimental Data and Fit Setup}\label{sec:ExpFit}

For our strong coupling determination we use the available $e^+e^-$ data ranging from 35 GeV to 207 GeV, which amounts to a total of 700 data points. We employ the minimal overlap model to treat correlations among systematic uncertainties accompanying the experimental bins. This gives rise to an experimental covariance matrix that is of the form
\begin{equation}
    \sigma_{ij}^{\rm exp} = 
    \delta_{ij} (\Delta_i^{\rm stat})^2 +
\delta_{D_iD_j}    {\rm min}(\Delta_i^{\rm sys},\Delta_j^{\rm sys})^2\,.
\label{sigmaexp}
\end{equation}
For the theoretical prediction, we first assess the uncertainties through renormalization scale variation, which in our setup corresponds to different choices of the profile parameters. To be definite, we take 5000 different sets of profile choices in our flat random scan and then determine the average $\bar x_i= (x_i^{\text{max}} + x_i^{\text{min}})/2$, as well as the uncertainty $\Delta^{\rm theo}_i = (x_i^{\text{max}} - x_i^{\text{min}})/2$ for each experimental bin $i$. Here, $x_i^{\text{max}}$ and $x_i^{\text{min}}$ are the boundaries of the interval obtained from varying the 17 theory parameters. In the following, we compute the correlation coefficient among different bins
\begin{align}\label{eq:theo-cor}
r_{ij}^{\text{theo}} = \frac{\left<(x_i - \bar x_i) (x_j - \bar x_j)\right>}
{\sqrt{\left<(x_i - \bar x_i)^2\right>}\sqrt{\left<(x_j - \bar x_j)^2\right>}}\,,
\end{align}
with $\left< y \right>$ being the average of the quantity $y$. Rescaling $r_{ij}$ by the corresponding uncertainties then gives rise to the theory covariance matrix
\begin{align}\label{eq:theo-cov}
\sigma_{ij}^{\rm theo} = \Delta_i^{\rm theo}\,\Delta_j^{\rm theo}\,r_{ij}^{\rm theo}\,.
\end{align}
The total covariance matrix entering the $\chi^2$ we minimize for the fit is then given by the sum of the experimental and theoretical covariance matrices:
$\sigma_{ij}^{\rm tot} = \sigma_{ij}^{\rm theo} + \sigma_{ij}^{\rm exp}$. Thus, the $\chi^2$ reads
\begin{align}\label{eq:chi2}
\chi^2 = \sum_{i,j=1}^{N_{\rm bins}} (\bar x_i - x_i^{\rm exp})\,
(\bar x_j - x_j^{\rm exp})\,(\sigma_{\rm tot}^{-1})_{ij}\,.
\end{align}
For our final results, we minimize Eq.~(\ref{eq:chi2}) w.r.t. $(\alpha_s,\Omega_1^{\rho},\Theta_1)$ and profile over the individual parameters in order to get the corresponding $1-\sigma$ uncertainties.

\section{Results}\label{sec:results}

\begin{figure}[t!]
\begin{minipage}{0.49\linewidth}
\centerline{\includegraphics[width=0.99\linewidth]{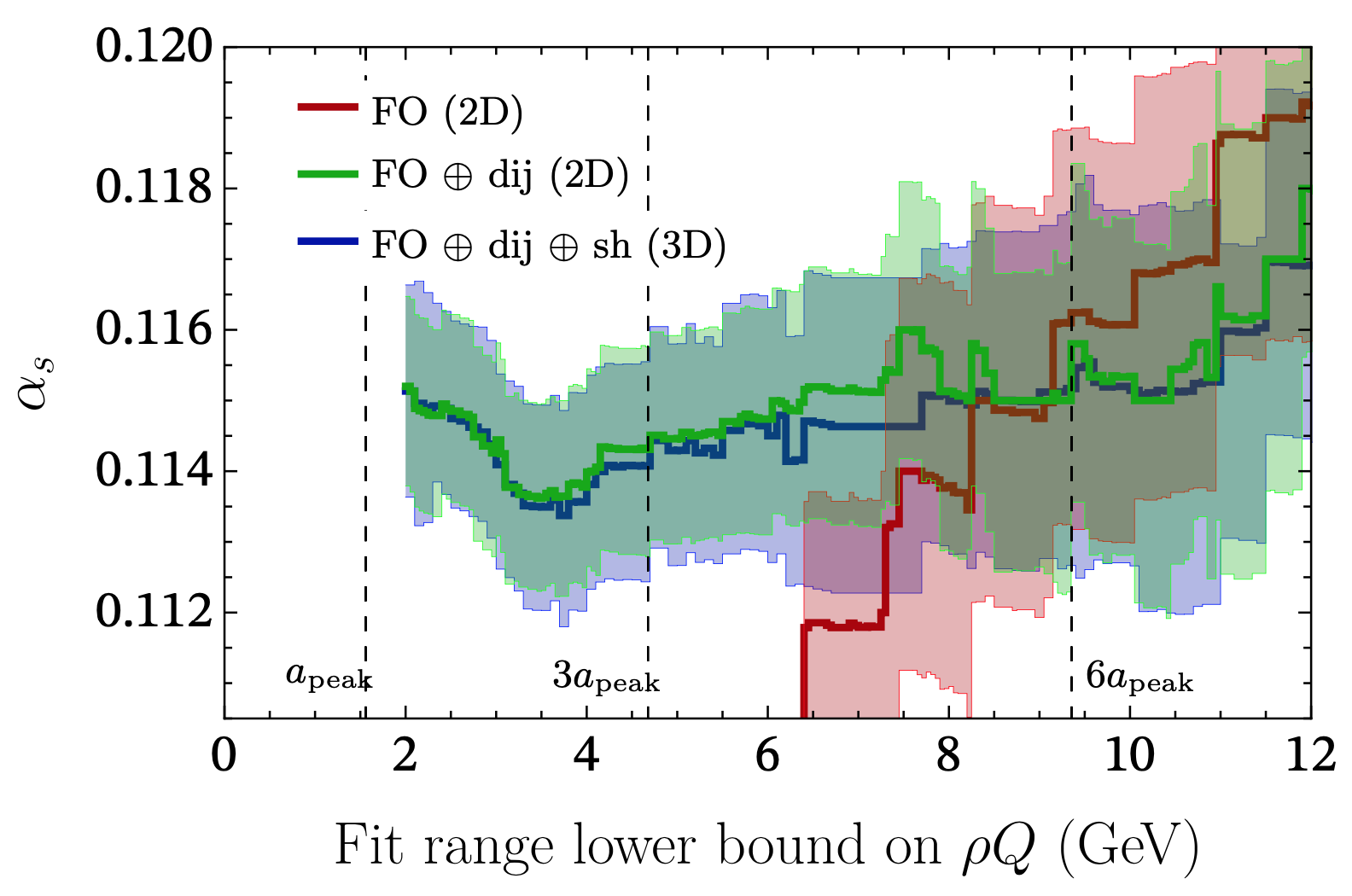}}
\end{minipage}
\hfill
\begin{minipage}{0.49\linewidth}
\centerline{\includegraphics[width=0.98\linewidth]{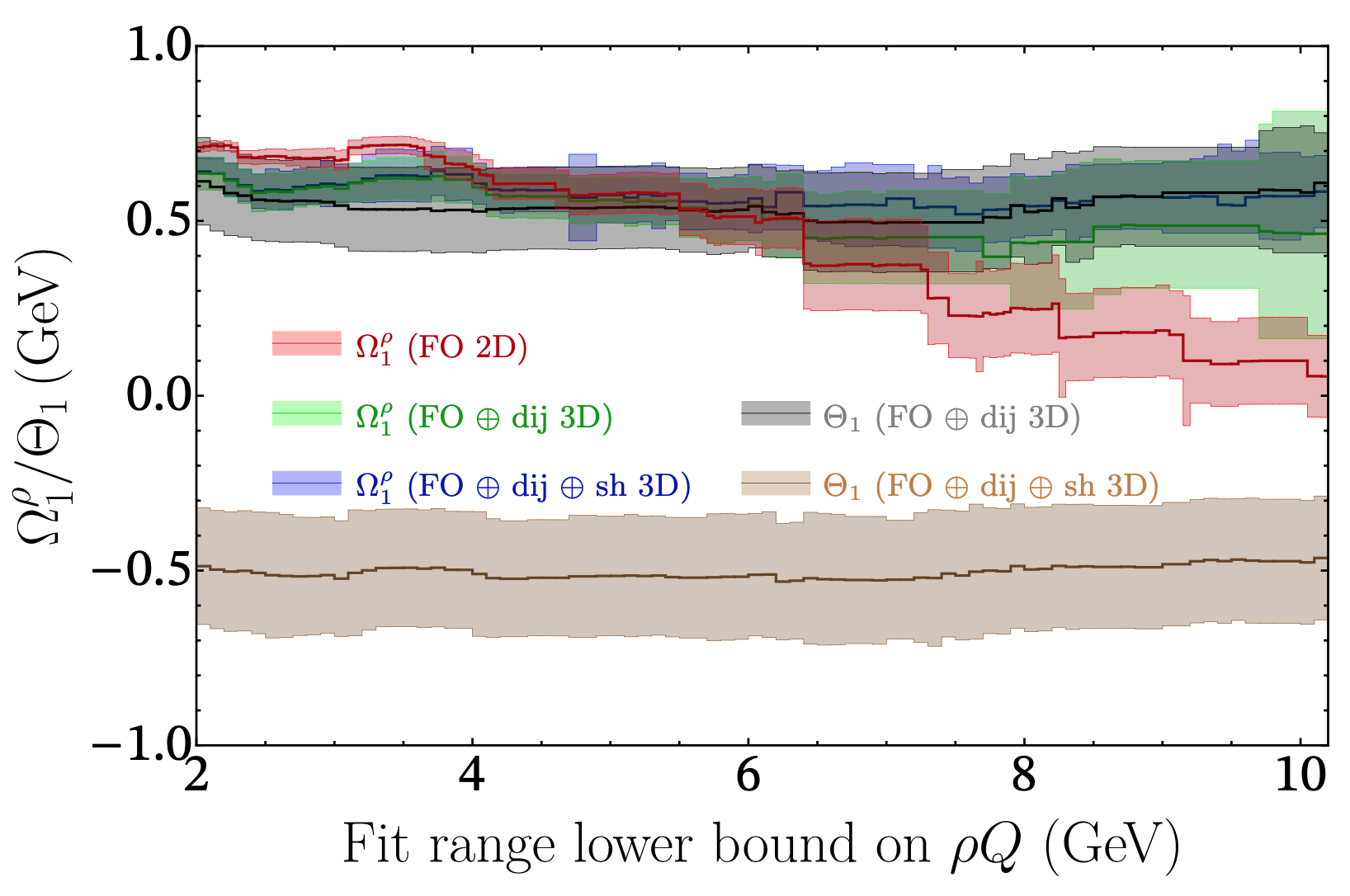}}
\end{minipage}
\caption[]{Results for $(\alpha_s,\Omega_1^{\rho},\Theta_1)$ as a function of the lower bound of the fit range for different levels of theoretical sophistication. The solid lines correspond to the central prediction for the given fit ranges. The bands denote the accompanying $1-\sigma$ uncertainties obtained from the fit.}
\label{fig:lowerBoundPlots}
\end{figure}
\begin{table}[t!]
    \centering
    \caption{Fit results for $\alpha_s,\Omega_1^{\rho}$ and $\Theta_1$ for different levels of theoretical sophistication. The central values are obtained from a weighted average over variations of the lower bound of the fit range. The same variation gives rise to the fit range uncertainty. The prescription used to obtain both of these results is discussed in Ref.~\protect\cite{Benitez:2025vsp}.
    \label{tab:fits}}
    \vspace{0.4cm}
    \resizebox{\textwidth}{!}{
    \begin{tabular}{|c|c|c|c|c|c|c||c|c|}
    \hline
Model & $\alpha_s(m_Z)$ &  th+exp  
& %NP err  
$\Omega_1^{\rho}$ & %NP err
$\Theta_1$ &  fit range %err 
& ~$\chi^2$/dof~ & $\Omega_1^{\rho}$\,[GeV] & $\Theta_1$\,[GeV] \\
\hline
Fixed Order 2D & ~$0.1166 \pm 0.0034$~ & ~$\pm\,0.0014$~ & ~$\pm\,0.0027$~ & -- & ~$\pm\,0.0015$~ & $1.108$  & $0.06 \pm 0.13$ & -- \\
FO\,+\,dijet 2D & $0.1148 \pm 0.0018$ & $\pm\,0.0010$ & $\pm\,0.0014$ & -- &$\pm\,0.0004$ & $1.055$ & $0.53 \pm 0.09$ & -- \\
FO\,+\,dijet 3D & $0.1156 \pm 0.0024$ & $\pm\,0.0010$ & $\pm\,0.0021$ & ~$\pm\,0.0004$~ & $\pm\,0.0007$ & $1.052$ & $\,0.52 \pm 0.08\,$ & $~\,\,0.53 \pm 0.13$ \\
\,FO\,+\,dijet\,+\,shoulder 3D\, & $0.1145 \pm 0.0020$ & $\pm\,0.0009$ & $\pm\,0.0018$ & $\pm\,0.0001$ &$\pm\,0.0003$ & $1.043$ & $0.57 \pm 0.09$ & \,$-0.50 \pm 0.17$\,  \\
        \hline
    \end{tabular}
    }
\end{table}
In Fig.~\ref{fig:lowerBoundPlots} we show our results for $(\alpha_s,\Omega_1^{\rho},\Theta_1)$ as a function of the lower bound of the fit range for different levels of theoretical sophistication. The corresponding numerical values are given in Table~\ref{tab:fits}. The left panel in Fig.~\ref{fig:lowerBoundPlots} displays the results for $\alpha_s(m_Z)$, whereas the right panel shows the corresponding results for the two non-perturbative parameters $\Omega_1^{\rho}$ and $\Theta_1$. The fixed-order prediction shown in red exhibits a linear behavior in terms of its dependence on the lower bound of the fit range. Thus, the extracted value for the strong coupling constant strongly depends on the choice of the fit range. This implies that a pure fixed-order prediction is not suitable to extract a reliable value for $\alpha_s(m_Z)$ without an arbitrary choice of the fit range. The results including dijet resummation, both for the setup with two fit parameters $(\alpha_s,\Omega_1^{\rho})$ denoted as (2D), as well as for the setup with three fit parameters $(\alpha_s,\Omega_1^{\rho},\Theta_1)$ denoted as (3D), are shown in green for $(\alpha_s,\Omega_1^{\rho})$ and gray for $\Theta_1$. Note that for these curves we do not yet include shoulder resummation. It is evident that dijet resummation provides a more robust setup, which enables a reliable determination of the strong coupling constant, as its value is remarkably insensitive to the choice of the fit range. This is also manifested in the small fit range uncertainty quoted in Table~\ref{tab:fits}. Another important aspect of this setup is the fact that the non-perturbative parameters $(\Omega_1^{\rho},\Theta_1)$ both turn out to be positive. Our best fit setup is finally shown in blue. As can be seen on the left panel in Fig.~\ref{fig:lowerBoundPlots}, including shoulder resummation does not impact the qualitative outcome for $\alpha_s(m_Z)$. However, for the non-perturbative parameter $\Theta_1$ we observe a change of sign. Interestingly, if we take the ratio between $\Omega_1^{\rho}$ and $\Theta_1$, which is conceptually related to the non-perturbative model function $\zeta(\rho)$ derived in Refs.~\cite{Caola:2021kzt,Caola:2022vea}, we find that our fit result agrees, within the quoted uncertainties, with their prediction for this 3-jet power correction parameter\footnote{I thank Silvia Ferrario Ravasio for pointing this out.}. We emphasize, however, that in our setup data favors a negative power correction in the 3-jet region of the HJM distribution only if shoulder resummation is included.

\section{Conclusions}\label{sec:Conclusions}

We have presented a thorough analysis on the extraction of the strong coupling constant from available $e^+e^-$ data, which we fit together with two non-perturbative parameters $\Omega_1^{\rho}$ and $\Theta_1$. In addition to a robust theoretical framework, provided by resummation in the dijet region, we further include the resummation of logarithms related to the left-Sudakov shoulder around the symmetric 3-jet configuration in the HJM spectrum. Correlations among parameters approximating the effects of missing higher orders are included by means of a theory covariance matrix, which is part of the $\chi^2$ we minimize in the fit. We obtain a value of the strong coupling constant that is compatible with similar determinations from more inclusive event shapes like Thrust~\cite{Abbate:2010xh,Benitez:2024nav} or the C-parameter~\cite{Hoang:2015hka} and further found evidence for a negative power correction in the 3-jet region only when including shoulder resummation in our setup.

\section*{Acknowledgments}

The work presented in these proceedings has been carried out in collaboration with A. Bhattacharya, A. H. Hoang, V. Mateu, M. D. Schwartz, I. W. Stewart and XY. Zhang. I am supported in part by the Spanish MECD grant No.\ PID2022-141910NB-I00, the EU STRONG-2020 project under Program No.\ H2020-INFRAIA-2018-1, Grant Agreement No.\ 824093, as well as by a JCyL scholarship funded by the regional government of Castilla y Le\'on and European Social Fund, 2022 call.

\end{document}